\documentclass[article]{emulateapj}
\slugcomment{Draft}

\shorttitle{Shear and Torsional Alfv\'en Waves}
\shortauthors{Laming}

\begin{document}

\title{The First Ionization Potential Effect from the Ponderomotive Force: On the
Polarization and Coronal Origin of the Alfv\'en Waves}


\author{J. Martin Laming\altaffilmark1}


\altaffiltext{1}{Space Science Division, Naval Research Laboratory, Code 7684, Washington DC 20375
\email{laming@nrl.navy.mil}}

\begin{abstract}
We investigate in more detail the origin of chromospheric Alfv\'en waves
that give rise to the separation of ions and neutrals, the First Ionization
Potential Effect (FIP), through the action of the ponderomotive force. In
open field regions, we model the dependence of fractionation on the plasma
upflow velocity through the chromosphere for both shear (or planar) and
torsional Alfv\'en waves of photospheric origin. These differ mainly
through their parametric coupling to slow mode waves. Shear Alfv\'en waves
appear to reproduce observed fractionations for a wider range of model
parameters, and present less of a ``fine-tuning'' problem than do torsional
waves. In closed field regions, we study the fractionations produced by
Alfv\'en waves with photospheric and coronal origins. Waves with a coronal
origin, at or close to resonance with the coronal loop, offer a
significantly better match to observed abundances than do photospheric
waves, with shear and torsional waves in such a case giving essentially
indistinguishable fractionations. Such coronal waves are likely the result
of a nanoflare coronal heating mechanism, that as well as heating coronal
plasmas releases Alfv\'en waves that can travel down to loop footpoints and
cause FIP fractionation through the ponderomotive force as they reflect
from the chromosphere back into the corona.

\end{abstract}

\section{Introduction}
The First Ionization Potential (FIP) Effect is the by now well known
abundance anomaly in the solar corona and slow speed solar wind, whereby
elements such as Fe, Si, and Mg with FIP less than about 10 eV, i.e. those
elements that are predominantly ionized in the solar chromosphere, are
enhanced in abundance in the solar corona with respect to solar photospheric
values by a factor of about three. High FIP elements are generally unchanged,
although He and Ne can also show abundance depletions. First discovered by
\citet{pottasch63}, a compelling explanation in terms of the ponderomotive
force arising as Alfv\'en waves propagate through or reflect from the
chromosphere has been advanced \citep{laming04,laming09,laming12,laming15}.

The ponderomotive force, discussed in more detail in section 2 and in the
Appendix, arises from the combined effects of the reflection and refraction
of Alfv\'en waves on the plasma ions. Plasma neutrals are not affected, since
the waves are fundamentally oscillations of the magnetic field. The time
averaged ponderomotive force is generally directed upwards in the solar
chromosphere, giving rise to an enhancement of ions over neutrals. In special
conditions, a downwards pointing force can arise instead, depleting the ions
to give rise to an ``Inverse FIP Effect'', as seen recently near a sunspot
\citep{doschek15}, and in the coronae of stars of later spectral type than
the Sun \citep{wood12}.

The time varying part of the same ponderomotive force is also responsible for
the excitation of acoustic waves by a parametric instability
\citep[e.g.][]{hollweg71}, and these acoustic waves can inhibit the FIP
fractionation. Consequently while the explanation of the FIP Effect in terms
of the ponderomotive force may be considered ``compelling'', much scope
remains for subtle details of the fractionation to be influenced by details
of wave-wave interactions in the chromosphere, and by the precise origin of
the Alfv\'en waves. Such topics will be a great interest for the forthcoming
solar missions Solar Orbiter \citep{muller13} and the Parker Solar Probe
\citep{fox16}.

The traditional view is that Alfv\'en waves, presumably resulting ultimately
from the mode conversion of p-modes \citep[see e.g.][]{cally16}, or from
other interactions of photospheric acoustic energy with magnetic flux tubes
\citep[see discussion in][]{cranmer05} are injected into a coronal loop
through the chromospheric footpoints from below. Unless a coincidental match
between the wave frequency and the loop resonance exists, significant
reflection of waves back down to the photosphere occurs
\citep{hollweg84,vanballegooijen11}. Alternatively, waves generated in the
corona will naturally be on resonance \citep[e.g.][]{ruderman02,dahlburg16},
at least until chromospheric evaporation or other activity changes the
coronal Alfv\'en speed or loop length. Such waves reflect from the top of the
chromosphere back into the corona. One of the conclusions of this paper will
be that the observed FIP Effect strongly favors these resonant waves, and
hence a coronal origin for the Alfv\'en waves.

Further, differences may also occur between shear (or planar) and torsional
Alfv\'en waves. Historically shear Alfv\'en waves, appropriate to an
unstructured solar atmosphere, have received most attention, though recent
observations have indicated the presence of torsional waves in spicules
\citep{depontieu12,sekse13} and in other regions of the chromosphere and
transition region \citep{jess09,depontieu14}. Cally (2017; see also
Mathioudakis et al. 2013; Jess et al. 2013) give a thorough review of solar
observations and interpretations in terms of the various manifestations of
planar or torsional Alfv\'en(ic) waves. While these different waves obey the
same Alfv\'en wave transport equations \citep[see
e.g.][]{heinemann80,khabib97,cranmer05}, they have different parametric
couplings to acoustic waves, and potentially different FIP fractionation may
result.

In this paper we investigate the effect of differing origins and
polarizations of Alfv\'en waves on the FIP fractionation pattern. Section 2
treats the coupling of torsional and shear Alfv\'en waves to slow mode waves,
including the effects of an upward plasma flow. The fractionation is also
subtly changed by this upward flow. Section 3 gives fractionation results for
the open magnetic field of a coronal hole, while section 4 gives the
corresponding results for a closed coronal loop. Section 5 concludes, with
some technical details concerning derivations of the expression for the
ponderomotive force and equations governing the torsional Alfv\'en waves
given in Appendices.

\section{Coupling to Slow Mode Waves}

\subsection{Torsional Alfv\'en Waves}
Torsional Alfv\'en waves are rotational oscillations of a cylindrical plasma
structure, of higher density than its surroundings. To keep the gas pressure
perturbation zero as in an Alfv\'en wave, an incompressible displacement must
be directed perpendicularly to the pressure gradient. Thus in a cylindrical
plasma column with untwisted magnetic field, the incompressible waves are
torsional oscillations \citep{vasheghani10}.
\citet{vasheghani10,vasheghani11} treat an ideal case with a discontinuity in
gas pressure and density at the periphery of the (assumed circular) flux
tube, whereas the possibly more familiar planar Alfv\'en wave applies in the
case of infinite and uniform plasma. Obviously a continuum of cases will
exist between these two idealizations, depending on the density or pressure
gradients in the background plasma. However in what follows, we will restrict
ourselves to comparing between these two limits.

The coupling of torsional Alfv\'en waves to slow mode waves is treated by
\citet{vasheghani11}, for the axisymmetric ($m=0$) case, extending as far as
linear terms in compressible variables. We follow their treatment and extend
to arbitrary $m$, where the azimuthal variation of the perturbation is given
by $\exp im\phi $, and include a background density gradient as seen in the
simulations of \citet{delzanna05} and included in the treatment of shear
Alfv\'en waves in \citet{laming12}, given by the usual hydrostatic
stratification of the chromosphere. The corona is expected and observed to
exhibit a filamentary structure \citep[e.g.][]{dahlburg12,testa13}, where the
transverse filament dimension is comparable to the longitudinal density scale
height in the chromosphere, both of which are much smaller than the Alfv\'en
wave wavelength. Consequently the thin flux tube approximation is valid. The
linearized equation of motion is
\begin{equation}
\rho{\partial\delta {\bf v}\over\partial t}+\rho\left(\delta{\bf v}\cdot\nabla\delta {\bf v}\right)=
{\bf J}\times {\bf B}-\nabla\delta P -{\bf g}\delta\rho
\end{equation}
for perturbed velocity, pressure and density $\delta {\bf v}$, $\delta P$,
and $\delta\rho$ respectively, with other symbols having their usual
meanings. In the $z$-direction (in cylindrical coordinates, and neglecting
the displacement current, with {\bf B} and {\bf g} along {\bf z})
\begin{equation}
\rho{\partial\delta v_z\over\partial t}=-{1\over 4\pi}
\left(\delta B_r{\partial \delta B_r\over\partial z}+\delta B_{\phi}{\partial \delta B_{\phi}\over\partial z}\right)
-{\partial \delta P\over\partial z}-g\delta\rho
\end{equation}
We recognize the first term in brackets $\left(\cdots\right)$ on the right
hand side as the ponderomotive force. For standing waves, such as in coronal
loops, it can be rewritten as
\begin{eqnarray}
\nonumber &&-{\partial\left(\delta
B_r^2+\delta B_{\phi}^2\right)\over 8\pi\partial z}={\partial\left(\epsilon
-1\right)\left(\delta E_r^2+\delta E_{\phi}^2\right)\over 8\pi\partial z}\\
&=&\Sigma _j{m_jc^2\over 2}{\partial\left(\delta E_r^2+\delta
E_{\phi}^2\right)/B^2\over \partial z},
\end{eqnarray}
where $\epsilon -1 = \Sigma _k\omega _{pk}^2/\Omega _k^2 = c^2/V_A^2=\Sigma
_j4\pi m_jc^2/B^2$ with $\Sigma _k$ being a sum over over particle species
$k$ with plasma frequency $\omega _{pk}$ and cyclotron frequency $\Omega _k$,
and $\Sigma _j$ being a sum over over particles $j$ each with mass $m_j$ in
the volume in which the density is measured. This leads to an instantaneous
ponderomotive force on each particle of $m_jc^2{\partial\left(\delta
E_r^2/B^2+\delta E_{\phi}^2/B^2\right)/\partial z}/2$. In fact as shown in
the Appendix A, this last form for the ponderomotive force is more general
than the negative gradient of the magnetic energy, which only holds in the
WKB approximation, and can be derived from the more general form within this
approximation. Note the change of sign between the initial and final
expressions in equation 3, due to the relative phases of $\delta B$ and
$\delta E$ in the standing wave. This sign change was omitted in equation 14
of \citet{laming12}. To lowest order this sign change makes no difference,
but when higher order terms are included, it reduces slightly the resulting
slow mode wave amplitudes given in that paper.

In the $r$-direction \citet{vasheghani11} give (their equation 4)
\begin{eqnarray}
\nonumber
\left[P+{B\delta B_z\over 4\pi}\right]_{r=0}&-&{A\over 2\pi}\rho{\partial\over\partial t}
\left(\delta v_r\over r\right)-{AB\over 16\pi ^2}{\partial ^2 \delta B_z\over\partial z^2}\\
&=&P_{ext} +{AJ^2\over 8\pi ^2} -{A\Omega ^2\over 2\pi ^2},
\end{eqnarray}
where $J=\delta B_{\phi}/r$, $\Omega = \delta v_{\phi}/r$ and $A$ is the area
of the flux tube cross section. While $J$ and $\Omega$ should be generalized
for $m>0$ for our purposes, in fact ultimately these terms tend to zero in
the thin flux tube approximation, leaving
\begin{equation}
{\partial\delta P\over\partial t}=-{B\over 4\pi}{\partial\delta B_z\over\partial t}
={B^2\over 4\pi}\left({\delta v_r\over r}+{m\delta v_{\phi}\over r}\right)\cos m\phi,
\end{equation}
where we have substituted for $\partial\delta B_z/\partial t$ from the
induction equation;
\begin{equation}
{\partial\delta {\bf B}\over\partial t}=\nabla\times\left(\delta {\bf v}\times {\bf B}\right),
\end{equation}
expanded following equation B1 in Appendix B below. Substituting for
$\left(\delta v_r+m\delta v_{\phi}\right)/r$ from the continuity equation;
\begin{equation}
{\partial\delta\rho\over\partial t}+\rho\nabla\cdot{\bf v}={\partial\delta\rho\over\partial t}
+\rho{\partial \delta v_z\over\partial z}+\rho\left({\delta v_r\over r}+
{m\delta v_{\phi}\over r}\right)\cos m\phi=0,
\end{equation}
into equation 5 yields
\begin{equation}
{\partial\delta P\over\partial t}={-B_z^2\over 4\pi\rho}
\left(i\omega\delta\rho+\rho{\partial\delta v_z\over\partial z}\right)
\end{equation}
with $\delta\rho\propto\exp\left(i\omega _st\right)$. Differentiating
equation 8 again with respect to $t$, and substituting from equation 2, where
$c_s$ is the sound speed, $\delta P =\delta\rho c_s^2$, and
$V_A=B_z/\sqrt{4\pi\rho }$ is the Alfv\'en speed, gives
\begin{eqnarray}
{c_s^2\over V_A^2}{\partial ^2\delta\rho\over\partial t^2}&=&\omega _s^2\delta\rho
+{\partial ^2\delta P\over\partial z^2}+g{\partial\delta\rho\over\partial z}\\
\nonumber &-& {\partial\over\partial z}\left\{{\rho c^2\over 2}{\partial\over\partial z}
\left(\delta E_r^2+\delta E_{\phi}^2\over B_z^2\right)\right\} +{\partial\rho\over\partial z}
{\partial\delta v_z\over\partial t}.
\end{eqnarray}
which with $\partial\rho /\partial z=\rho /L$, $i\omega _s\delta\rho
=-\left(ik_s+1/L\right)\rho\delta v_z$, $\partial\delta\rho /\partial
z=\left(ik_s+1/L\right)\delta\rho$ and $\partial\delta v_z/\partial
z=ik_s\delta v_z$ \citep{laming12} becomes
\begin{eqnarray}
\nonumber\delta\rho &&\left\{\omega _s^2\left(c_s^2+V_A^2\right) + c_s^2V_A^2\left(ik_s+1/L\right)^2 +
V_A^2g\left(ik_s+1/L\right)\right\}\\
=&&V_A^2{\partial\over\partial z}\left\{{\rho c^2\over 2}{\partial\over\partial z}
\left(\delta E_r^2+\delta E_{\phi}^2\over B_z^2\right)\right\}-{i\omega\rho V_A^2\over L}\delta v_z.
\end{eqnarray}
The first term in curly brackets on the left hand side, $\omega _s^2c_s^2$,
is new and does not appear in the treatment of shear Alfv\'en waves (see
below, subsection 2.4), and leads to different properties of the coupling of
torsional Alfv\'en waves to slow modes \citep[e.g.][]{vasheghani11}.
Identifying
\begin{eqnarray}
\nonumber\omega _s&=&2\omega _A/n\\
\nonumber \Re k_s&=&\sqrt{\left(2\Re k_A/n\right)^2-g^2/4c_s^4}\\
\nonumber \Im k_s&=&-g/2c_s^2\\
L&=&-c_s^2/g
\end{eqnarray}
where $n$ is an integer, \citep[see e.g.][]{landau76}, we find
\begin{equation}
\omega _s^2\delta\rho ={\partial\over\partial z}\left\{{\rho c^2\over 2}{\partial\over\partial z}
\left(\delta E_r^2+\delta E_{\phi}^2\over B_z^2\right)\right\}-{i\omega\rho\over L}\delta v_z,
\end{equation}
where the new term $\omega _s^2c_s^2=4\omega _A^2c_s^2/n^2$ cancels exactly
with the term in $4k_A^2V_A^2c_s^2/n^2$. Thus unlike the case of shear
Alfv\'en waves, resonant generation of sound waves when $V_A=c_s$ does not
occur with torsional Alfv\'en waves, as in \citet{vasheghani11}. Integrating
by parts, equation 12 can be rewritten as
\begin{eqnarray}
\nonumber\delta v_z&=&-{ic^2\over 2\omega _s}\left(1+ik_sL\over 2+ik_sL\right)
{\partial\over\partial z}\left(\delta E_r^2+\delta E_{\phi}^2\over B_z^2\right)\\
&=&-{i\over 2\omega _s}\left(1+ik_sL\over 2+ik_sL\right)
{\partial\over\partial z}\left(\delta v_r^2+\delta v_{\phi}^2\right)
\end{eqnarray}
since $\delta {\bf E}=\delta {\bf v}\times{\bf B}/c$. The parametric
generation of sound waves by torsional Alfv\'en waves is lower than for shear
Alfv\'en waves because the possibility of a resonance does not exist.

\subsection{Inclusion of Plasma Flow}
We introduce a steady plasma flow velocity ${\bf U} = U{\bf \hat{z}}$, where
${\bf \hat{z}}$ is a unit vector in the z-direction and $\partial U/\partial
t=0$ into equations 1 and 7. We assume $U << c_s$ for the time being, so that
the background chromospheric density profile is not changed. Equation 12
becomes
\begin{eqnarray}
 & & \left(\omega _s^2+{i\omega _sU\over L}\right)\delta\rho
={\partial\over\partial z}{\rho c^2\over 2}{\partial\over\partial z}
\left(\delta E_r^2+\delta E_{\phi}^2\over B_z^2\right)\\
\nonumber &-&{\partial\over\partial z}\left\{\rho U\delta v_z\left(ik_s-{1\over L}\right)-{\delta\rho U^2\over L}
\right\}-{i\omega _s\rho\over L}\delta v_z,
\end{eqnarray}
which after substituting $i\omega _s\delta\rho
=-\left(ik_s+1/L\right)\rho\delta v_z$ and integrating by parts gives
\begin{eqnarray}
\nonumber & & \left\{\left(\omega _s+{iU\over L}\right)^2+{\omega _s\over k_s}{U\over L^2}\left(1+k_s^2L^2\right)
-{ik_sU^2\over L}+{\omega _s^2\over 1+ik_sL}\right\}\delta v_z\\
& & =-i{\omega _sc^2\over 2}{\partial\over\partial z}\left(\delta E_r^2+\delta E_{\phi}^2\over B_z^2\right).
\end{eqnarray}
In steep density gradients, $k_sL << 1$ and the second term in brackets
$\left[ ... \right]$ on the left hand side can dominate if $U>0$, reducing
$\delta v_z$.

\subsection{Back Reaction of Slow Modes on Alfv\'en Waves}
The term on the right hand side of equation 15 is unpacked as follows
\begin{eqnarray}
-i{\omega _sc^2\over 2}{\partial\over\partial z}\left(\delta E^2\over B_z^2\right)&=&
-i{\omega _s\over 2}{\partial\over\partial z}\left(\delta B^2\over 4\pi\left(\rho+\delta\rho
\right)\right)\\
\nonumber &\simeq &-i{\omega _s\over 2}{\partial\over\partial z}\left({\delta B^2\over 4\pi\rho}\left[1-
{\delta\rho\over\rho}+ ...\right]\right)\\
\nonumber \simeq \omega _sk_A\delta V_A^2 &+&\delta v_z\delta V_A^2\left(k_Ak_s-{k_s^2\over 2}
-i{k_A\over L}+i{k_s\over 2L}\right),
\end{eqnarray}
where $\delta V_A^2=\delta B^2/4\pi\rho $ and $\partial\delta V_A^2/\partial
z=2ik_A\delta V_A^2$, and $i\omega _s\delta\rho =
-\left(ik_s+1/L\right)\rho\delta v_z$ as before. The extra terms in $\delta
v_z$ have the effect of reducing the slow mode wave amplitude in the regions
of steep density gradient, where the Alfv\'en wave becomes evanescent
\citep[see e.g.][]{rakowski12}. Ignoring for the time being the flow
velocity,
\begin{eqnarray}
\delta v_z&=&{\omega _sk_A\delta V_A^2\over\omega _s^2-\delta V_A^2
\left(k_Ak_s-{k_s^2/2}-i{k_A/L}+i{k_s/2L}\right)}\\
\nonumber &\rightarrow & {k_A\delta V_A^2\over\omega _s}\quad{\rm where}~ \Re k_A\rightarrow 0,
~ \Re k_s\rightarrow 0\\
\nonumber &\rightarrow & {\omega _sk_A\delta V_A^2\over\omega _s^2-
\left(1-i/2\Re k_AL+1/4\Re k_AL\right)\delta V_A^2g^2/16c_s^4}
\end{eqnarray}
otherwise, after putting $\Im k_A=1/4L$, $\Im k_s=1/2L$, and $\left(\Re
k_s\right)^2=4\left(\Re k_A\right)^2-g^2/4c_s^2$. Equations 17 indeed give
lower $\delta v_z$ when $\Re k_A=0$.

\subsection{Shear Alfv\'en Waves Revisited}
We give here the corresponding equation for shear Alfv\'en waves,
\begin{eqnarray}
\nonumber & & \biggl\{\left(\omega _s+{iU\over L}\right)^2+
\omega _sk_sU -{i\omega _sU\over L}-{ik_sU^2\over L}\\
\nonumber & & -k_s^2c_s^2-ik_sg-\delta V_A^2\left(k_Ak_s-{k_s^2\over 2}
-i{k_A\over L}+i{k_s\over 2L}\right)\biggr\}\delta v_z\\
& & =-i{\omega _sc^2\over 2}{\partial\over\partial z}
\left(\delta E_r^2+\delta E_{\phi}^2\over B_z^2\right).
\end{eqnarray}

This is slightly different to equation 14 from \citet{laming12} to the same
order in $\delta v_z$ for an assumed isothermal chromosphere ($\gamma =1)$,
(and with the sign of the ponderomotive force corrected), although numerical
results are very similar. The reason for the difference is that we have taken
a linear relation between $\delta\rho$ and $\delta v_z$, $i\omega
_s\delta\rho =-\left(ik_s+1/L\right)\rho\delta v_z$, whereas \citet{laming12}
took higher orders into account.

\subsection{Slow Mode Waves from Convection}
The different couplings of shear and torsional Alfv\'en waves to slow mode
waves are the features that give rise to the subtly different fractionation
patterns the waves produce, as will be shown below. Another source of slow
mode wave derives from convection below the photosphere. \citet{heggland11}
model this process, resulting in upgoing acoustic waves in the chromosphere,
with an amplitude approximately constant with altitude of order 7.0 km
s$^{-1}$ \citep[see also Figure~3ab in][]{vieytes05,carlsson15,kato11} in the
region where strong wave transmission upwards from the photosphere occurs.
The amplitude growth that would be expected as the wave travels upwards in
progressively decreasing density of the chromosphere is presumably balanced
by energy losses to radiation and thermal conduction. This wave amplitude is
added in quadrature to the amplitude of parametrically generated slow mode
waves for use in model calculations. If the resulting amplitude is greater
than the local sound speed, such that a shock would develop, all
fractionation is switched off. The rationale for this is that the
discontinuity associated with a shock front will cause cause reflection of
upgoing sound waves once overtaken by the shock. Downgoing and upgoing waves
can interact and cause a cascade to microscopic scales. The mixing connected
with this cascade will inhibit all further fractionation.

\subsection{Model Calculations}
We model Alfv\'en wave propagation on a model loop anchored in the
chromosphere at each end, largely following the extensive description given
in \citet{laming15}. The chromospheric model is taken from \citet{avrett08},
which gives the plasma density, electron temperature, and ionization fraction
of hydrogen as a function of altitude above the photosphere. Ionization
fractions for other elements are calculated using atomic collisional data
tabulated in \citet{mazzotta98}, \citet{bryans09}, \citet{nikolic13} and
\citet{kingdon96}. Photoionization rates are evaluated using coronal spectra
incident from above based on \citet{vernazza78} or more modern calculations
\citet[e.g.][]{huba05}, with photoionization cross sections from
\citet{verner96}.

Alfv\'en wave propagation is treated in a chromospheric magnetic field model
given by \citet{athay81}, typically with an expansion of a factor of 5
between the photosphere and corona, where the field is straight and uniform
(the coronal loop is taken to be straightened out with a chromosphere at each
end). The Alfv\'en wave transport equations are given by \citet{cranmer05}
and \citet{laming15}, and are
\begin{equation}
{\partial I_{\pm}\over\partial t}+\left(u\mp V_A\right){\partial I_{\pm}\over\partial z}
=\left(u\pm V_A\right)\left({I_{\pm}\over 4L} +{I_{\mp}\over 2L_A}\right)
\end{equation}
where ${\bf I}_{\pm}=\delta {\bf v}\pm \delta {\bf B}/\sqrt{4\pi\rho}$ are
the Els\"asser variables representing waves propagating in the $\mp$
z-directions.

Given the Alfv\'en wave profile, the ponderomotive force is calculated from
equations A2 or A7, divided by a further factor of two to account for the
average force in terms of the peak electric field fluctuation $\delta E$. The
chromospheric model employed here is not specifically a loop footpoint
chromosphere, and so any effect of the coronal Alfv\'en waves on the
chromospheric structure must be accounted for. This is treated by
\citet{laming15}, who rewrites the equation for a gravitationally stratified
hydrostatic equilibrium, including the ponderomotive force, which in turn
depends on the resulting density gradient. This modification begins to set in
at ponderomotive accelerations around $10^6$ cm s$^{-2}$ and ultimately leads
to saturation. It makes a small difference to the fractionations calculated
here, but is included because it also allows fractionations relative to
hydrogen to be computed \citep[instead of just oxygen as was done until now,
e.g.][]{laming15}. This modified expression for the ponderomotive
acceleration is
\begin{equation}
a={a_0\over 1+\left(\xi _h/4\right)\left(\nu _{eff}/\nu _{hi}\right)
\left(\delta v^2/v_h^2\right)}
\end{equation}
where $\nu _{eff}=\nu _{hi}\nu _{hn}/\left(\xi _h\nu _{hn}+\left(1-\xi
_h\right)\nu _{hi}\right)$ is the effective collision frequency of element
$h$ (in this case hydrogen) in terms of its collision frequencies when
ionized $\nu _{hi}$, and when neutral $\nu _{hn}$, and $v_h^2=k_{\rm B}T/m_h
+v_{sm}^2+v_{\mu turb}^2$ is the square of the hydrogen speed, in terms of
its thermal speed, the amplitude of slow mode waves propagating through the
chromosphere, and the amplitude of microturbulence in the chromospheric
model. Since the ponderomotive force separates ions from neutrals, its effect
on the background plasma to smooth out density gradients depends on the
coupling between ionized and neutral hydrogen, and is strongest in regions
where hydrogen is fully ionized ($\xi =1$). This modification to the density
scale length is also applied to $L$ in equation 11.

\begin{figure*}[t]
\centerline{\includegraphics[scale=0.9]{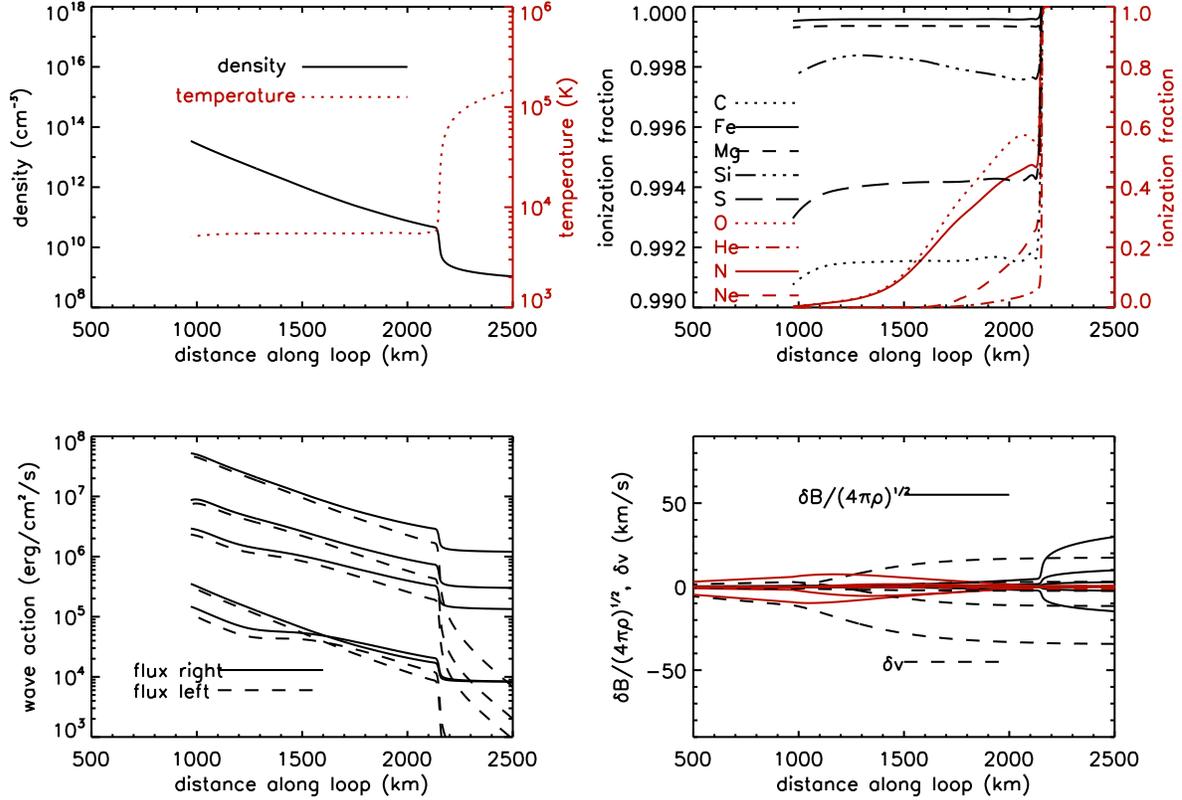}}
\caption{\label{fig1}The chromospheric portion of the open field model. Top left shows the
density and temperature structure of the chromosphere. Top right shows chromospheric
ionization fractions for selected elements, low FIPs in black to be read on the
left hand axis, and high FIPs in red to be read on the right.
Bottom right show the variation of the components of the Els\"asser
variables for the five wave frequencies considered, real parts in black and imaginary
parts in red, and bottom left shows the
corresponding wave energy fluxes.}
\end{figure*}

\begin{figure*}[t]
\centerline{\includegraphics[scale=0.9]{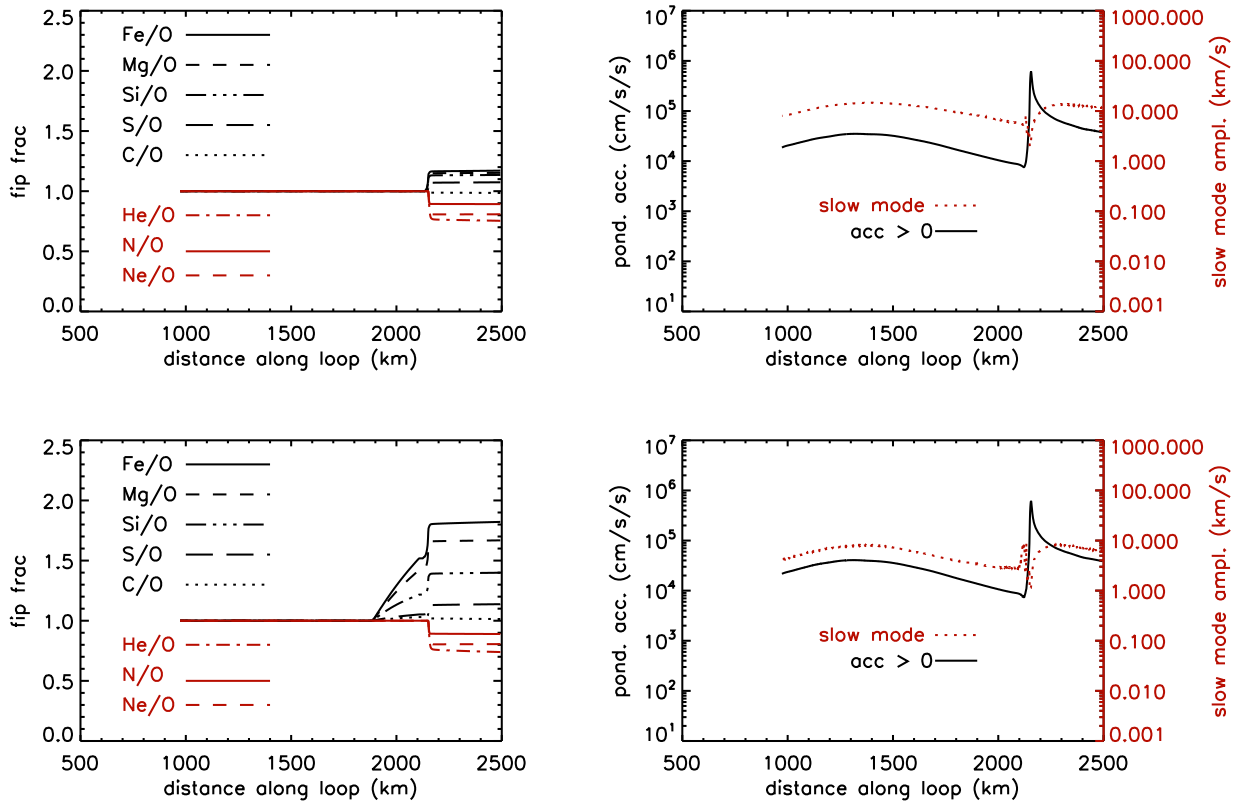}}
\caption{FIP fractionation for selected elements (left) and ponderomotive acceleration
and associated slow mode wave amplitude, in red, and to be read on the right hand axis
(right). The top panels show results for shear
Alfv\'en waves, the lower ones for $m=0$ torsional Alfv\'en waves. The flow velocity is
taken to be 0.1 km s$^{-1}$ at a chromospheric density of $10^{10}$ cm$^{-3}$, and is
insignificantly different from the case with zero flow velocity. \label{fig2a}}
\end{figure*}

\begin{figure*}[t]
\centerline{\includegraphics[scale=0.5]{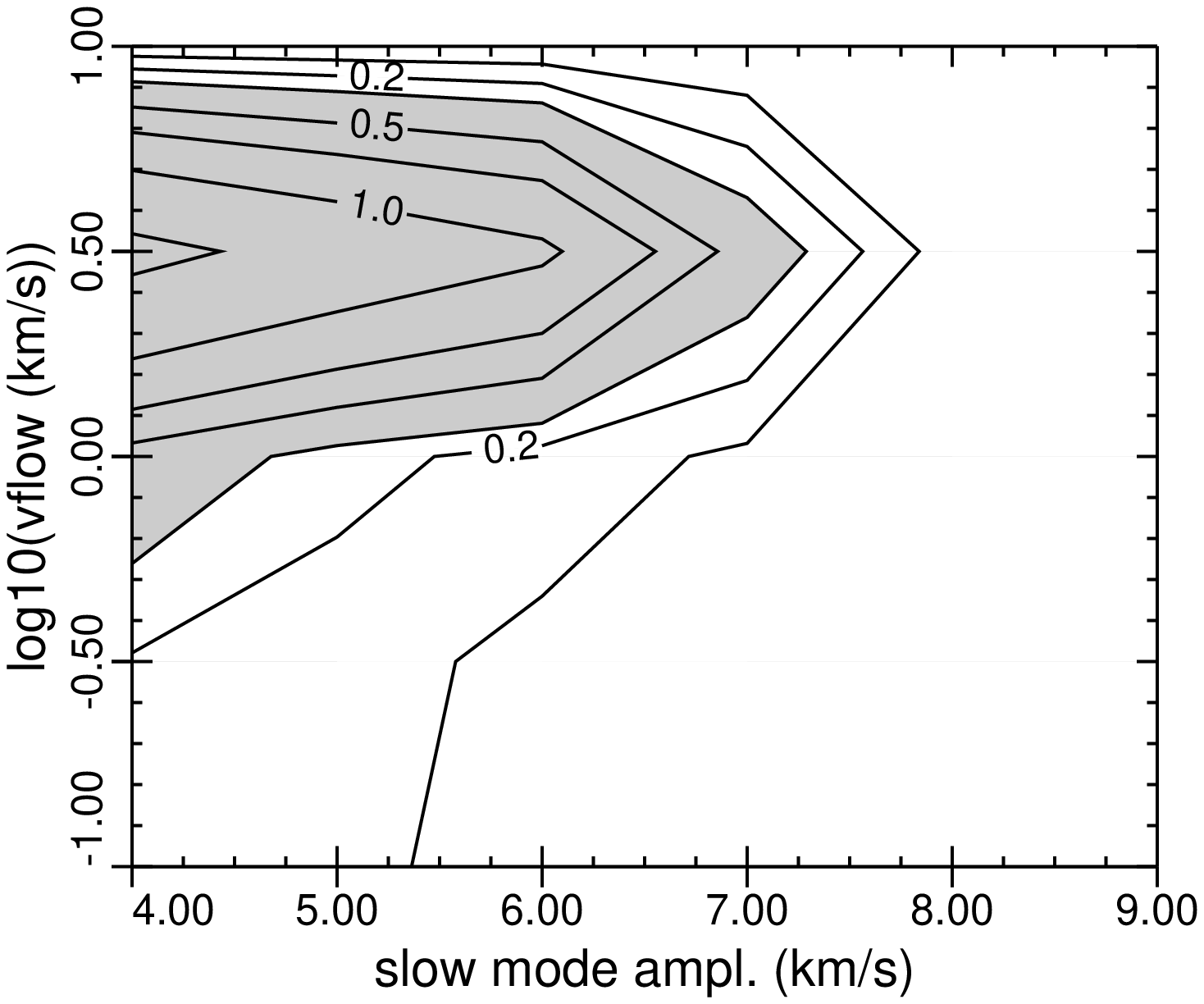}\includegraphics[scale=0.5]{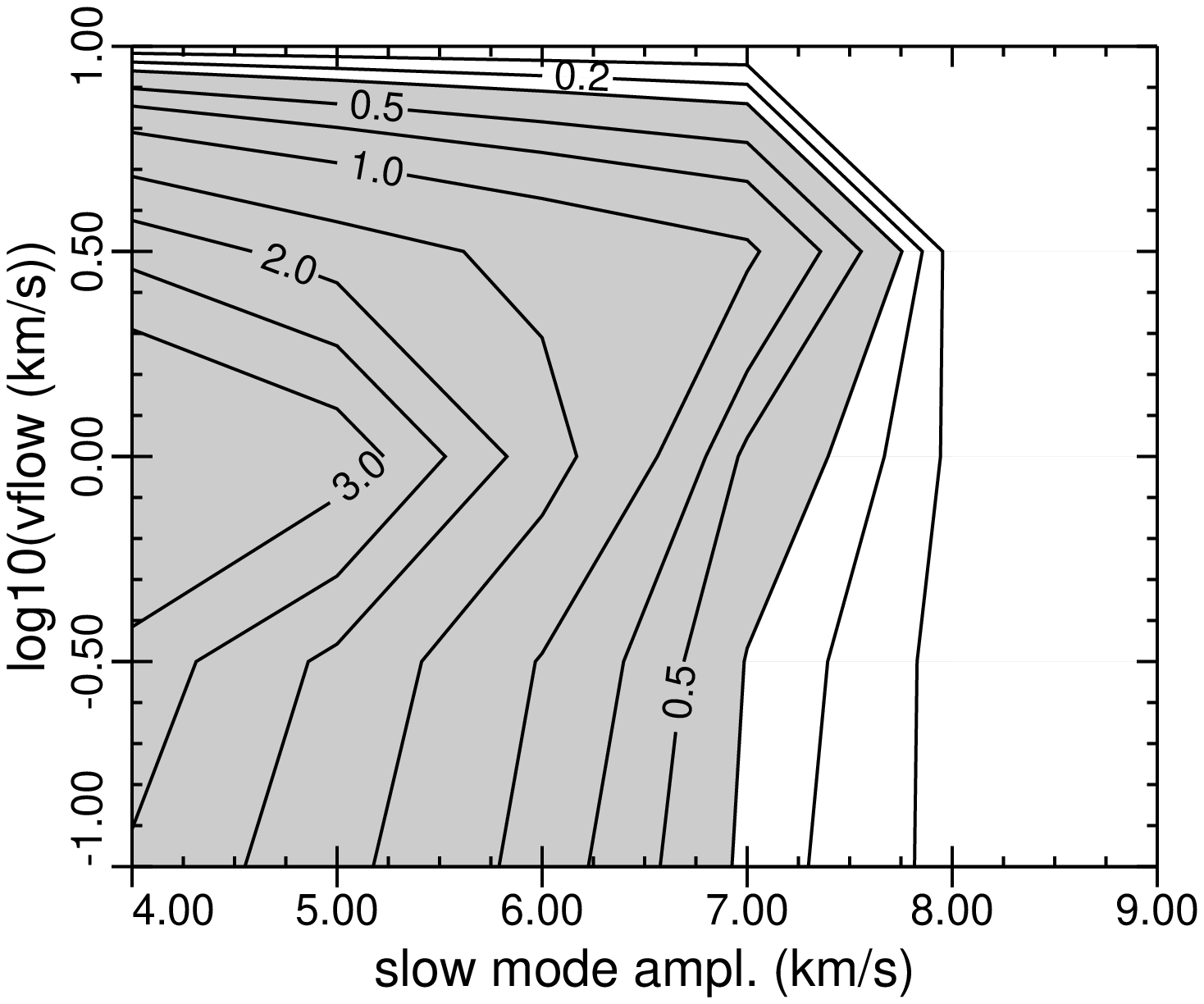}}
\vskip 0.5 truein
\caption{Sensitivity of the Fe/O fractionation presented as logarithmic contours in the
upflow velocity - slow mode amplitude plane. Left hand panel shows shear Alfv\'en
waves, right hand panel shows torsional Alfv\'en waves. Fractionations greater than
2 that violate observational values are shaded gray. \label{fig2b}}
\end{figure*}

With these definitions, the fractionation by the ponderomotive acceleration
for element $k$ (with ions and neutrals of element $k$ denoted by $ki$ and
$kn$ respectively) is given by solving the momentum equations
\begin{eqnarray}
\nonumber \rho _{ki}u_{ki}{\partial u_{ki}\over\partial z}+{\partial P_{ki}\over\partial z}
&=&-\rho_{ki}g-\rho_{ki}\nu_{ki}\left(u_{ki}-u\right) +\rho_{ki}a \\
\rho _{kn}u_{kn}{\partial u_{kn}\over\partial z}+{\partial P_{kn}\over\partial z}
&=&-\rho_{kn}g-\rho_{kn}\nu_{kn}\left(u_{kn}-u\right).
\end{eqnarray}
These are the same momentum equations as given elsewhere
\citep[e.g.][]{laming15}, but with inertial terms added on the left hand side
to account for the plasma flow. Following \citet{schwadron99} we form
$\nu_{kn}\partial P_{ki}/\partial z + \nu_{ki}\partial P_{kn}/\partial z$ to
find
\begin{equation}
{\partial\over\partial z}\left(P_k + \rho _ku_k^2\right)={\nu _{eff}\xi _ka\rho _k
\over \nu _{ki}}
\end{equation}
where $P_{ki}=\xi P_k$, $P_{kn}=\left(1-\xi\right)P_k$,
$\rho_{ki}=\xi\rho_k$, $\rho _{kn}=\left(1-\xi\right)\rho _k$, and we have
approximated
\begin{eqnarray}
\nonumber\rho _k\nu _{eff}\left({\xi _k\over \nu _{ki}}u_{ki}{\partial u_{ki}\over\partial z}
+{1-\xi _k\over\nu _{kn}}u_{kn}{\partial u_{kn}\over\partial z}\right)&\simeq &
\rho_ku_k{\partial u_k\over\partial z}\\
&=&{\partial\rho u_k^2\over\partial z}
\end{eqnarray}
since $\rho _ku_k=$ constant in the final step. We put $P_k=\rho
_k\left(kT/m_k +\Sigma v_{||}^2/2\right)$ where the terms in brackets are the
ion thermal speed and a quadrature sum over all nonthermal longitudinal
velocity oscillations, slow mode waves, microturbulence, etc. The
fractionation is evaluated from equation 22 by splitting the integration up
into pieces where $kT/m_k +\Sigma v_{||}^2/2 +u_k^2$ is approximately
constant and calculating
\begin{equation}
{\rho _k\left(z_u\right)\over\rho _k\left(z_l\right)}=\exp\left\{
\int _{z_l}^{z_u}{2\xi _ka\nu _{eff}\over \nu _{ki}\left(2kT/m_k+\Sigma v_{||}^2+2u_k^2
\right)}dz\right\}.
\end{equation}
in each piece,  where $u_k$ is specified to give zero fractionation in the
absence of $a$, and where the constant of integration also as been chosen to
keep $\rho_k\left(z_u\right) =\rho_k\left(z_l\right)$ when $a=0$. The total
fractionation is then the product of all the individual pieces.

\section{Open Magnetic Field}
We first demonstrate the effects of shear and torsional Alfv\'en waves in
open magnetic field. The basic chromospheric model and Alfv\'en wave
propagation are illustrated in Fig. 1, and follow very closely the set up in
\citet{laming12,laming15}. The magnetic field is taken from
\citet{banaszkiewicz98} and matched to the chromospheric magnetic field taken
from \citet{athay81}. The density profile is specified to match that in
equation 3 of \citet{cranmer05}. One departure from previous work is that we
now include an outward flow velocity. This does not affect the Alfv\'en wave
propagation much, since everywhere the flow is significantly sub-Alfv\'enic,
but as will be seen below does have an effect on the generation of slow mode
waves in the regions of strong density gradient.

Equations 19 are integrated starting from 500,000 km altitude back to the
chromosphere. At the coronal start point, the outgoing Alfv\'en waves
dominate \citep{cranmer05}. Five waves are simulated, with frequencies and
amplitudes chosen to match Fig. 3 in \citet{cranmer07} and for example Figs.
9 or 15 in \citet{cranmer05}. The four panels of Fig. 1 show the
chromospheric portion of the solution, with the density and temperature
structure of the model chromosphere (top left), and ionization balance
profiles for various important elements (top right; low FIP elements in black
to be read on the left hand $y$-axis, high FIP elements in red to be read on
the right hand $y$-axis), the left and right going wave energy fluxes (bottom
left) and the profiles of the components of the Els\"asser variables (bottom
right).

Figure 2 shows the elemental fractionations (left panels) and the profiles of
ponderomotive acceleration and slow mode wave amplitude (right panels). The
top half shows the results for a shear Alfv\'en wave, the bottom for the
torsional Alfv\'en wave. The difference in fractionation arises from the
different coupling to the slow mode waves. We have assumed an upward flow
speed of 0.1 km s$^{-1}$ where the chromospheric density is $10^{10}$
cm$^{-3}$ (see below for discussion and justification), which gives
essentially the same results for velocity put equal to zero. At an altitude
of 1200 km, the slow mode wave energy density is about 1 erg cm$^{-3}$, which
if the wave is moving in one direction only corresponds to a flux of about
$5\times 10^5$ ergs cm$^{-2}$s$^{-1}$. \citet{cranmer15} consider acoustic
waves fluxes of this magnitude (and larger) as a result of ponderomotive
driving by chromospheric Alfv\'en waves, although they do not compute the
coupling explicitly. \citet{arber16} and \citet{brady16} do compute this
coupling, and also find ponderomotive accelerations and slow mode wave
amplitudes broadly consistent with Fig. 2. These authors however do not see
the reduction in slow mode wave amplitude in the region of steep density
gradient according to equation 17 above, presumably because they model
propagating slow mode waves generated deeper in the chromosphere.

\begin{table*}
\begin{center}

\caption{Open Field Fractionations. }
\bigskip

\begin{tabular*}{\textwidth}{|c @{\extracolsep{\fill}} |cccccc|ccc|cc|c|}
\hline
Ratio & \multicolumn{3}{c}{shear} & \multicolumn{3}{c}{torsional} &
\multicolumn{3}{c}{ACE}& \multicolumn{2}{c}{Ulysses}& UVCS\\
      & 0 &0.1  & 1.0 km s$^{-1}$ & 0 & 0.1  & 1.0 km s$^{-1}$ & Solar Max. & Genesis & Solar Min & PCH& CH& \\
\hline  
He/O& 0.75 & 0.75 & 0.74 &0.74& 0.74&  0.71 &  0.51 & 0.48 & 0.45 &  & & 0.45-0.55 \\
C/O & 0.97 & 0.99 & 0.98 &1.00& 1.02&  1.04 &  1.11 & 1.15 & 1.16 & 1.22& 1.23& 0.9-1.1\\
N/O & 0.88 & 0.89 & 0.88 &0.88& 0.89&  0.87 &       &      &      & 1.11 & 1.11& \\
Ne/O& 0.80 & 0.81 & 0.79 &0.80& 0.81&  0.79 &  0.57 & 0.60 & 0.67 & 0.63& 0.63& 0.3-0.4\\
Mg/O& 1.13 & 1.15 & 1.18 &1.57& 1.67&  2.40 &  1.89 & 1.61 & 1.37 & 1.55& 1.54& 0.95-2.45\\
Si/O& 1.11 & 1.14 & 1.16 &1.34& 1.40&  1.79 &  2.60 & 2.40 & 1.99 & 1.74& 1.80& 0.9-1.8\\
S/O & 1.06 & 1.07 & 1.08 &1.11& 1.14&  1.22 &       &      &      & 2.19& 2.35& \\
Fe/O& 1.14 & 1.17 & 1.21 &1.70& 1.82&  2.79 &  2.08 & 1.90 & 1.56 & 1.41& 1.43& 0.65-1.35\\
\hline
\end{tabular*}

\end{center}
\tablecomments{ACE observational results are from \citet{pilleri15}, Ulysses
results from \citet{vonsteiger16}, and UVCS from \citet{ko06}.} \label{tab1}
\end{table*}

With the upflow speed increased to 1.0 km s$^{-1}$ at a chromospheric density
of $10^{10}$ cm$^{-3}$, the slow mode wave amplitudes in the region of strong
fractionation are decreased by the addition of the terms in $U$ and $U^2$ in
equations 15 and 18, and the fractionation is increased. Interestingly there
is now a stronger difference between shear and torsional Alfv\'en waves, with
the latter producing stronger fractionation. These results are summarized in
Table 1, and compared to various observations. The ``ACE'' observations come
from analysis of reprocessed data from the Advanced Composition Explorer
(ACE) by \citet{pilleri15}, where we quote the coronal hole measurements for
solar maximum (1999-2001), solar minimum (2006-2009), and for measurements
during 2001-2004, the period of operation of the Genesis mission. The Ulysses
results for polar coronal holes (PCH) and low latitude coronal holes (CH)
come from a similar reprocessing of the Ulysses archive given in
\citet{vonsteiger16}, while the UVCS results come from remotely sensed
spectroscopic measurements with the UltraViolet Corona-Spectrograph (UVCS) on
SOHO reported by \citet{ko06}. Taking a proton flux in the fast wind at 1 AU
of $\sim 2\times 10^8$ cm$^{-2}$s$^{-1}$ translates to a flux at the solar
surface of $\sim 10^{13}$ cm$^{-2}$s$^{-1}$. At a density of $10^{10}$
cm$^{-3}$, the upward flow velocity implied is $10^3/f$ cm s$^{-1}$, where
$f$ is the area filling factor of the flow in the coronal hole, and the
speeds taken above correspond to $f=0.1$ and $f=0.01$.

The differences between shear and torsional Alfv\'en waves are explored
further in Fig 3. Contours of the Fe/O ratio are plotted in the parameter
space defined by the upward flow speed and the chromospheric slow mode wave
amplitude resulting from convection. Hitherto we have assumed an amplitude of
7 km s$^{-1}$, but here we consider values between 4 and 9 km s$^{-1}$. We
also consider a wider range of upflow speeds than above, tacitly assuming
that the upflow is driven by the ${\bf J}\times{\bf B}$ term in the momentum
equation, as would be the case for a Type II spicule, leaving the
chromospheric density profile unchanged. Contours for shear Alfv\'en waves
are plotted in Fig. 3a, and for torsional Alfv\'en waves in Fig. 3b. The
shaded regions correspond to Fe/O $> 2$, and is the region that would violate
most observations of fractionation in the fast wind. We can see that for
shear waves, observational constraints can be satisfied for a much wider
range of upflow speeds and slow mode wave amplitudes than is the case for
torsional waves, because shear wave driver generates more parametric slow
modes that stabilize the fractionation than does the torsional wave driver.
Since the slow mode wave amplitude in the chromosphere is inevitably
variable, \citep[e.g.][]{heggland11} we argue that shear Alfv\'en waves are
more plausible.

Such an inference is of some interest. \citet{cranmer05} describe how
photospheric sound waves can initiate kink oscillations in flux tubes that
travel up to the corona. Upon reaching the corona where the initially
isolated photospheric flux tubes merge, the kink oscillation becomes
Alfv\'enic, either Alfv\'en or fast mode (depending on magnetic geometry) and
parallel propagating. In a homogenous corona, these would be shear Alfv\'en
waves, whereas inhomogeneity in the coronal plasma, or vortical photospheric
motions might result in torsional coronal Alfv\'en waves. \citet{verth16},
following \citet{goossens14}, illustrate how a kink mode may mode convert to
a $m=1$ torsional Alfv\'en wave. The difference between shear and torsional
waves is not necessarily negligible. \citet{vasheghani12} argue that
torsional Alfv\'en wave undergo a slower parallel nonlinear cascade than do
shear Alfv\'en waves, which could have implications for coronal heating and
solar wind acceleration, although the effect is largest when sound and
Alfv\'en speeds are equal.

\citet{requerey15} observe the dynamics of quiet sun magnetic structures. In
such cases, torsional oscillations act to stabilize multi-threaded magnetic
filaments, whereas planar perturbations corresponding to the forcing by
granular motions tend to shred the structure. The observed lifetimes and
morphological changes of magnetic flux concentrations strongly suggest
perturbation by granular convection, and not the characteristic torsional
oscillations of a thin flux tube, again consistent with our inference above.
On the other hand \citet{stangalini17} see elliptically polarized kink
oscillations of chromospheric small-scale magnetic elements (SSMEs), which
would reduce slow mode wave excitation (circularly polarization would
suppress it completely).

Lastly, several authors \citep[e.g.][]{depontieu11} have in recent years
suggested that a significant fraction of the solar corona and wind could be
the result of chromspheric spicules, more specifically ``Type II Spicules",
that accelerate material straight out of the chromosphere at speeds up to 100
km s$^{-1}$ \citep{martinez13}. Evidence exists for both transverse and
torsional oscillations \citep{sekse13}, though \citet{raouafi16} suggest that
the torsional motion might be due to the unwinding of a flux rope rather than
an actual torsional wave. Observed amplitudes are comparable to those in the
open field model above. Figure 3 shows the fractionations expected for shear
and torsional waves, assuming the same wave spectrum as for the coronal hole
considered above. FIP fractionation essentially disappears at upflow speeds
comparable to the sound speed ($\sim 8$ km s$^{-1}$) and above, as seen for
example in polar jets by \citet{lee15}. We follow \citet{klimchuk14} and
\citet{lionello16} and argue therefore that Type II spicules cannot be a
significant source of plasma to supply the solar corona or slow speed solar
wind, because no means exists within such a model to provide the FIP
fractionation. All fractionated plasma must move up to the corona in a more
gradual fashion.

\begin{figure*}[t]
\centerline{\includegraphics[scale=0.9]{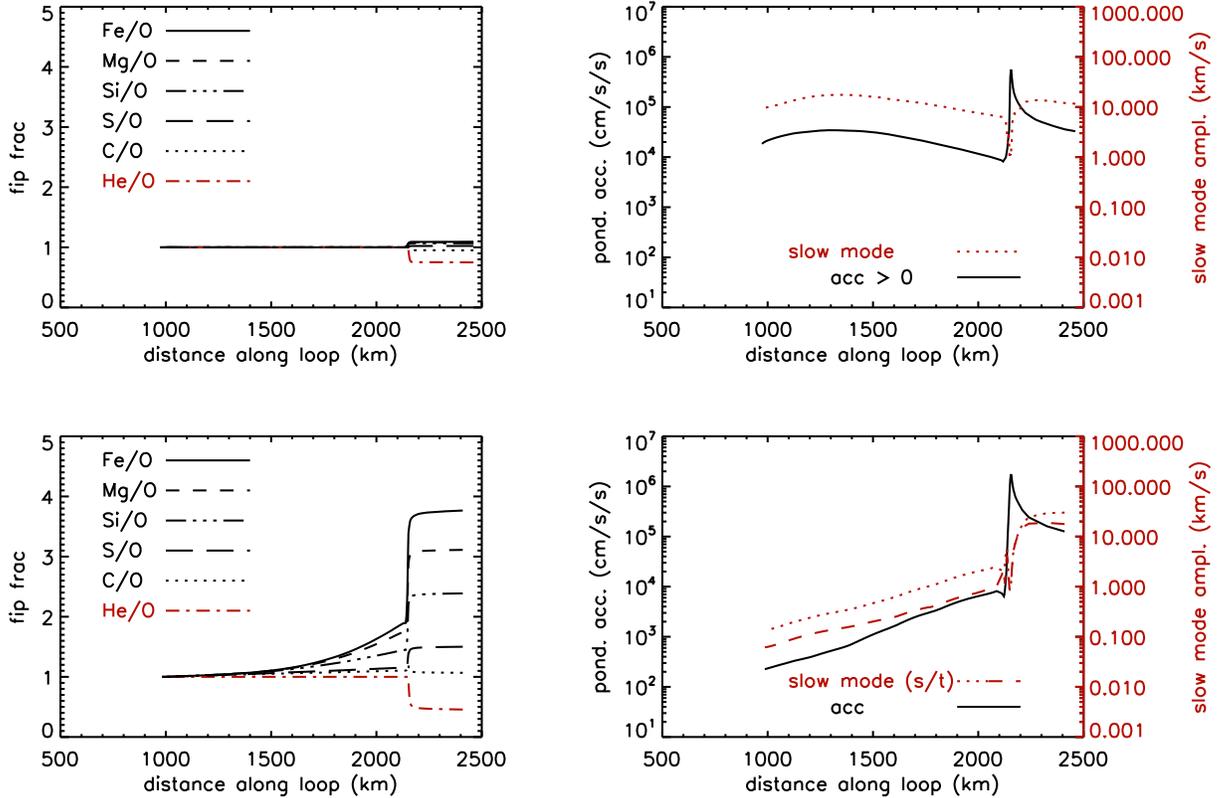}}
\caption{Fractionations (left) and ponderomotive acceleration/slow mode wave
amplitude (right) for a closed loop. The top panels show the case with similar
waves (assumed shear) injected upwards from one footpoint as in the open field case.
The bottom
panels show the case with one wave resonant with the loop, developing higher
ponderomotive acceleration and fractionation. Slow mode amplitudes are indicated for
both shear (dotted) and torsional (dashed) resonant waves.\label{fig3a}}
\end{figure*}

\begin{figure*}[t]
\centerline{\includegraphics[scale=0.9]{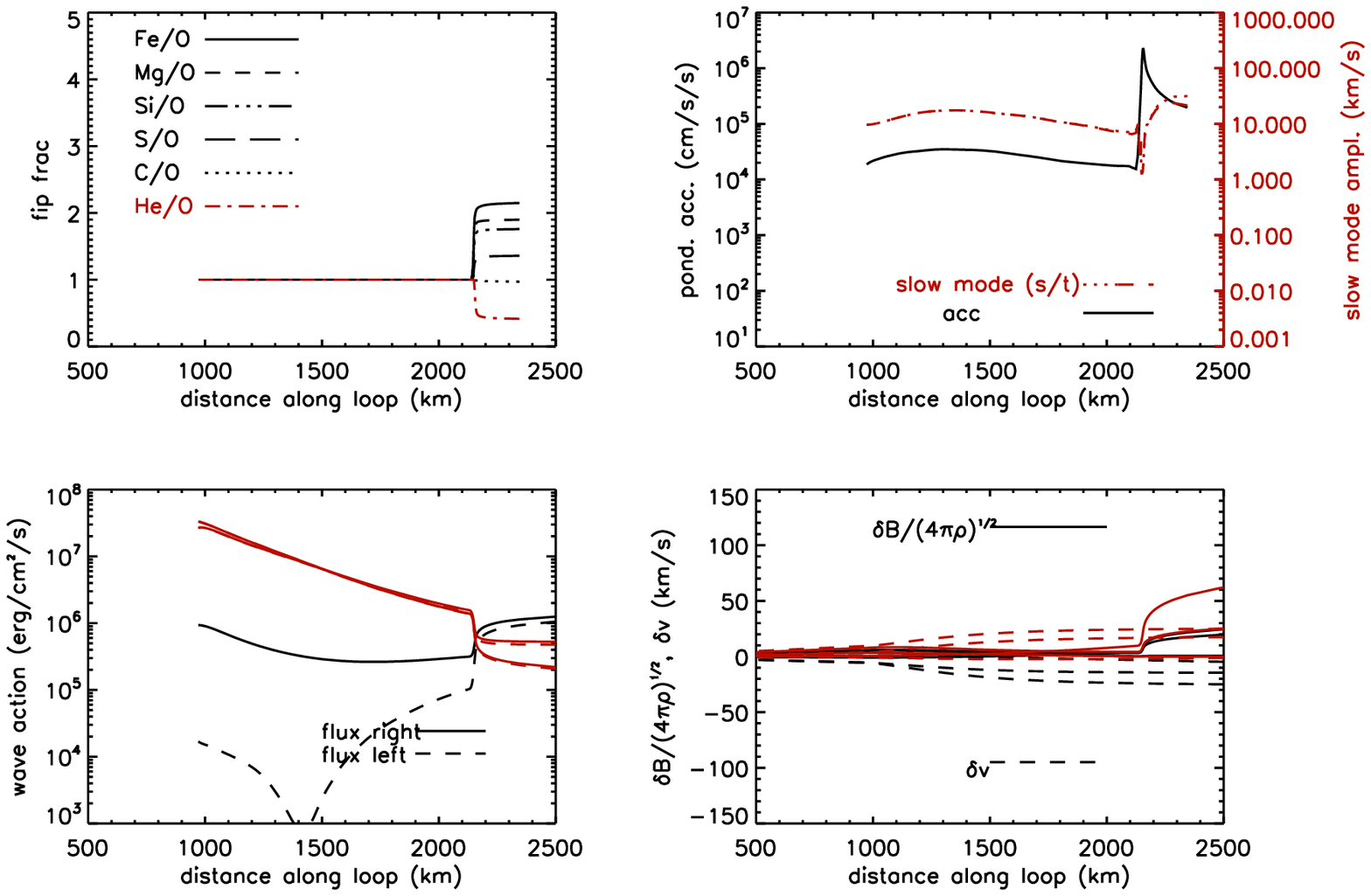}}
\caption{Top panels show fractionations (left) and ponderomotive acceleration/slow mode wave
amplitude (right) for a closed loop with both a resonant wave and three and five minute
waves injected from
the footpoint. Bottom panels show the wave energy fluxes (left), with the resonant wave in black
and the three and five minute photospheric waves in red, and components of the Els\"asser
variables (right), with real parts in black, imaginary parts
in red.\label{fig3b}}
\vskip 0.25 truein
\end{figure*}

\section{Closed Magnetic Loop}
We now consider a closed magnetic loop, and demonstrate the effect of a
resonance in the coronal magnetic structure.  The loop is taken to be 75,000
km long, with a coronal magnetic field of 10 G. The chromospheric upflow
speed at a density of $10^{10}$ cm$^{-3}$ is 0.1 km~s$^{-1}$. Although we do
not consider the coronal consequence of this flow, we presume it to be
associated with evaporation, one of the processes by which mass is supplied
to the corona. Figure \ref{fig3a} shows the fractionations (left) and
ponderomotive acceleration/slow mode wave amplitudes (right) for this case.
The top panels show a case with three minute and five minute \citep[see
e.g.][]{heggland11} Alfv\'en waves injected at the footpoint, here assumed to
be shear Alfv\'en waves, and with similar amplitudes to those in the open
field case. The fractionation is in all respects very similar to that in the
open field case with the same flow velocity of 0.1 km s$^{-1}$ (see Fig. 2,
and note the different vertical scale). He/O is depleted by the same amount,
$0.75$, and a similarly weak FIP fractionation is produced. The bottom panels
show a case with one resonant wave (taken to be a shear or torsional Alfv\'en
wave) with a frequency $\omega = V_A/L=0.0435$ rad s$^{-1}$ (the fundamental,
period = 2.4 minutes), presumed to have a coronal origin. The amplitude of
the coronal wave in this model is about 90 km s$^{-1}$, at a density of about
$7\times 10^8$ cm$^{-3}$ at the loop apex. Much stronger FIP fractionation
results, with a depletion of He/O by about 0.5.

Figure \ref{fig3b} shows in the top panel the fractionations and
ponderomotive acceleration/slow mode wave amplitudes for a model where both
sets of waves are present, i.e. the three and five minute Alfv\'en waves
propagating up from the footpoint and the coronal resonant wave. Here the
depletion of He/O is even stronger, at 0.4, but the FIP fraction is now
intermediate between the two cases above. This arises because the extra slow
mode waves produced by the upcoming Alfv\'en waves inhibit fractionation low
in the chromosphere, while the He depletion which occurs only at the top of
the chromosphere near the strong density gradient is relatively unaffected.
The lower two panels of Fig. \ref{fig3b} show the wave energy fluxes (left)
and the wave amplitudes $\delta v$ and $\delta B/\sqrt{4\pi\rho}$ (right). In
each case the resonant coronal wave is shown as a black line, while the
nonresonant waves from the footpoint are shown as red lines.

The results from Figs. 4 and 5 are summarized in Table 2, and compared with
various observations \citep{zurbuchen02,vonsteiger00,bryans09}. The model
chromosphere used here \citep{avrett08} connects with a fairly low density
corona (few $\times 10^8$ cm$^{-3}$), resulting in high predicted coronal
Alfv\'en wave amplitudes. Higher coronal densities require lower coronal
Alfv\'en wave amplitudes to produce the necessary FIP fractionation, with
wave amplitude going approximately as $1/\sqrt{\rho }$ for a standing wave.

\begin{table*}[t]
\begin{center}
\caption{FIP Fractionations in Closed Magnetic Field}
\begin{tabular*}{\textwidth}{|c @{\extracolsep{\fill}}|ccc|c|ccc|}
\tableline\tableline
&\multicolumn{3}{c}{models 75,000km}&100,000 km& \multicolumn{3}{c}{observations} \\
Ratio& bkgd & cor & bkgd+cor& bkgd& a& b& c\\
\tableline
He/O & 0.75& 0.46/0.44& 0.41/0.40& 0.21& 0.68-0.60& 0.29-0.75 & \\
C/O & 0.95& 1.07/1.07& 0.98/0.97& 0.99& 1.36-1.41& 1.06-1.37 & \\
N/O & 0.87& 0.74/0.74& 0.72/0.72& 0.55&  0.72-1.32& 0.22-0.89 & \\
Ne/O& 0.79& 0.60/0.60& 0.56/0.56& 0.36&  0.58& 0.38-0.75 & \\
Na/O& 1.10& 3.68/3.77 & 2.07/2.08& 5.48&  & & 7.8${+13\atop -5}$\\
Mg/O& 1.08& 3.11/3.31&  1.90/1.91& 4.38&  2.58-2.61& 1.08-2.36& 2.8${+2.3\atop -1.3}$\\
Al/O& 1.08& 2.91/3.08&  1.93/1.94& 4.45&  & & 3.6${+1.7\atop -1.2}$\\
Si/O& 1.07& 2.40/2.51&  1.77/1.79& 3.55& 2.49-3.11& 1.36-3.24& 4.9${+2.9\atop -1.8}$\\
S/O & 1.02& 1.51/1.54&  1.37/1.38& 1.88&  1.62-1.92& 1.23-2.68& 2.2$\pm 0.2$\\
K/O & 1.10& 3.94/4.30& 2.17/2.20& 6.11&  & &  1.8${+0.4\atop -0.6}$\\
Ca/O& 1.10& 3.90/4.26&  2.17/2.19& 6.13&  & & 3.5${+4.3\atop -1.9}$\\
Fe/O& 1.09& 3.79/4.13&  2.17/2.20& 6.00&  2.28-2.90& 0.96-2.46& 7.0${+1.4\atop -1.2}$\\
\tableline
\end{tabular*}
\end{center}
\tablecomments{From left to right, model FIP fractionations correspond
to those in Figs. 4 (top and bottom) and 5 (top). For each model, the entries are
that for three and five minute nonresonant waves (bkgd),
one resonant wave (cor), and the combination of one resonant and two nonresonant
waves (bkgd+cor), respectively.
Observational ratios are taken from, (a) Zurbuchen
et al. (2002), given relative to O, (b) von Steiger et al. (2000), relative to O, and
(c) Bryans et al. (2009), given
relative to the mean of O, Ne and Ar. Ranges quoted from von Steiger et al. (2000) include their uncertainties.}
\label{tab4}
\end{table*}

\begin{figure*}[t]
\centerline{\includegraphics[scale=0.5]{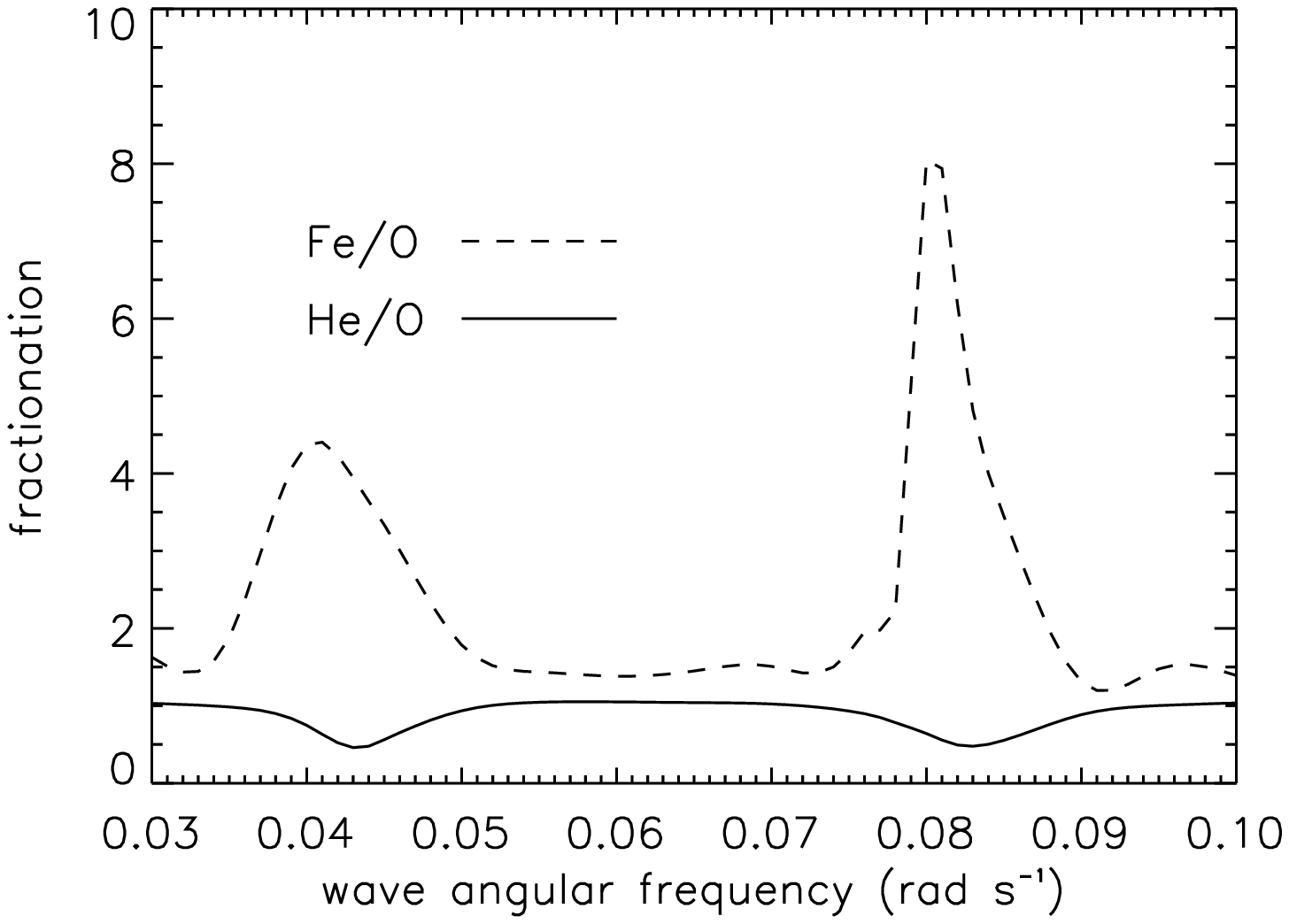}\includegraphics[scale=0.5]{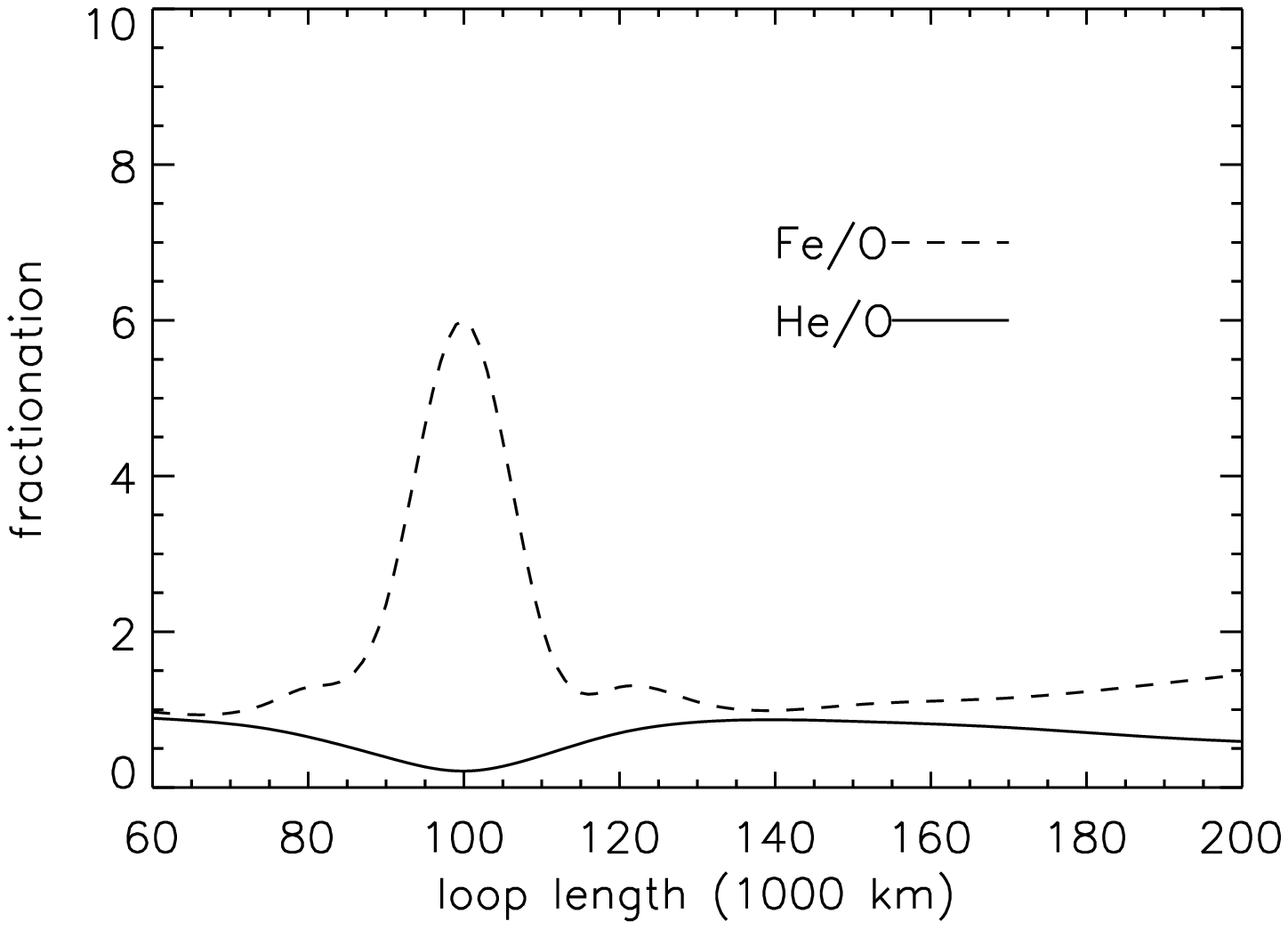}}
\caption{Left: Fractionations of Fe/O (dashed lines) and He/O (solid line) as a
function of coronal wave frequency for the model $10^5$ km loop. At the fundamental and first
harmonic the Fe/O maximizes and the He/O minimizes, coming closest to
observations at these points. Right: Same fractionations as a function of loop
length for 3 and 5 minute footpoint waves. The resonance at $10^5$ km corresponds
to 3 minute waves entering the loop. \label{fig6}}
\end{figure*}

\section{Discussion and Conclusions}
In order to reproduce FIP fractionation, or depletions of He/O beyond about
0.7, the presence of a resonant Alfv\'en wave appears necessary.
\citet{kasper12} and more recently \citet{kepko16} have identified slow speed
solar wind regimes where the He abundance (usually relative to H) can be as
low as 1\% - 2\%. \citet{rakowski12} found that resonant waves on longer
loops with weaker magnetic fields produced the most significant He
depletions. While FIP fractionation can come from coronal or photospheric
waves, the He depletion requires the ponderomotive acceleration to be
concentrated at the top of the chromosphere, where He is neutral but where O
and H are becoming ionized. This only happens if waves are impinging on the
chromosphere from the top, and being reflected back upwards again. This
probable necessity of resonant waves implies a coronal origin for the waves,
presumably as a result of nanoflares \citep[e.g.][]{dahlburg16}.

The importance of resonant waves is emphasized further in Fig. 6. The left
panel shows the fractionations of Fe/O and He/O as a function of wave
frequency for the model loop introduced above. The calculations are
normalized so the same total wave energy is present in the
chromosphere-loop-chromosphere structure, just distributed differently as the
resonant properties change. At the loop fundamental, (0.0435 rad s$^{-1}$)
and the first harmonic, the Fe/O fractionation increases to about 4-8, and
the He/O ratio decreases to about 0.5. In between these resonances, He/O is
close to being unfractionated, while a mild FIP effect appears in Fe/O,
comparable to open field values (see Table 1). Notice also that the
resonances in the fractionation are not sharp. They have a finite width of
order 0.01 rad s$^{-1}$, a significant fraction of the wave angular
frequencies themselves, since the waves reflect from a range of chromospheric
heights and not from a sharp boundary. These resonances would be even broader
if we allowed for coronal wave damping, by introducing an imaginary part in
the wave frequency. The important point about the width of these resonances
is that while resonance is required, such that a coronal wave origin is most
likely, it need not be precise. Changes in the loop resonant frequency due to
changes in magnetic field by reconnection, or changes in loop plasma density
due to chromospheric evaporation are unlikely to destroy the fractionation,
except arguably in flares.

The right panel of Fig. 6 shows a similar calculation but this time with just
the 3 and 5 minutes waves, with fractionation shown as a function of loop
length. An ``accidental'' resonance occurs at a loop length of 100,000 km,
where the three minute waves can become trapped in the loop. At this
location, the Fe/O fractionation increases and the He/O ratio decreases,
similarly to the resonances above. The fractionation pattern at 100,000 km is
given in the fifth column of Table 2. Since a large fraction of the wave
travel time from one footpoint to the other occurs inside the chromosphere at
each end, the corresponding resonance for the five minutes wave occurs not at
5/3 times 100,000 km but at a loop length longer than 200,000 km. The coronal
magnetic field in these examples is rather weak, at 10 G, and the loop
lengths required for resonance are rather long, so it appears unlikely than
an ``accidental'' resonance with photospheric three or five minute wave will
happen frequently in a realistic coronal loop. Instead, an explanation of
loop resonance in terms of waves generated within the coronal loop appears
more plausible.

Candidate processes for production of Alfv\'en waves in the solar corona are
Alfv\'en resonance \citep[e.g.][]{ruderman02} or nanoflares
\citep[e.g.][]{dahlburg16}. In the former case torsional $m=1$ Alfv\'en waves
result from the mode conversion of a kink oscillation at the plasma layer
within a loop where the Alfv\'en wave transit time from one loop footpoint to
the other matches the kink oscillation period \citep[e.g.][]{jess16}. In the
latter case the shear Alfv\'en waves could also be produced. The different
FIP fractionations produced by shear and torsional Alfv\'en waves appear to
be too subtle for a reliable diagnosis of the Alfv\'en wave polarization.
More promising seems to be the observations of strong non-thermal line
broadening spatially coincident with regions of strong FIP fractionation in
an active region observed by \citet{baker13}. The line broadening is
presumably due to longitudinal waves, being observed just above loop
footpoints on the solar disk. With this interpretation, shear Alfv\'en waves
are favored by the wave amplitudes required. Torsional Alfv\'en waves do not
produce strong enough slow modes to match the observations.

The preference for shear Alfv\'en waves generated in the corona suggests a
nanoflare heating model. Based partly on the simulations in Dahlburg et al.
(2016; see also Kigure et al. 2010), we argue that coronal nanoflares release
Alfv\'en waves and heat. The Alfv\'en waves cause FIP fractionation at loop
footpoints magnetic connected to the nanoflare site, followed by evaporation
as the electrons conduct the heat down to the chromosphere. In such a
scenario, the waves are likely to be in resonance with the coronal loop,
reproducing the fractionation as shown in Fig. 6. The reconnection of a
closed loop filled with plasma in such a manner with open field is most
likely the source of the slow speed solar wind
\citep[e.g.][]{lynch14,raouafi16}. Also such a scenario will require a
coronal Alfv\'en speed greater than the speed of electron thermal conduction.
It is possible that this requirement is easier to meet in longer loops, where
the thermal conduction front has to travel further to the loop footpoints and
can become more degraded with distance.

Much more work remains to be done on coronal abundance anomalies and the
associated wave physics responsible for them. This paper makes an attempt to
show how the physics of the fractionation places constraints on the mechanism
of coronal heating and mass supply in completely complementary ways to
traditional methods of observation and inference. The focus on the Alfv\'en
waves and their polarizations and resonance probes different properties of
the coronal heating mechanism, and in connection with {\it in situ}
observations may also be hoped to yield insight into the origin of the solar
wind in the forthcoming era of the Parker Solar Probe and Solar Orbiter.

\acknowledgements This work has been supported by NASA HSR grants (NNH16AC39I
and NNH13AV38I) and by Basic Research Funds of the CNR.

\appendix
\section{Alternative Routes to the Ponderomotive Force}
\citet{laming09,laming15} give derivations by writing down the Lagrangian for
a system of particles and waves, and using the known result for the
partitioning of energy between kinetic, magnetic and electric in an Alfv\'en
wave. In subsection 2.2 above, it is derived from the MHD momentum equation.
Curiously, the more fundamental form of the expression does not come directly
from the MHD equation, but requires further manipulation.

First, consider the force per unit volume, $f_z$ on the polarization and
magnetization induced by turbulence in the plasma
\begin{equation}
f_z=\delta {\bf P}\cdot{\partial \delta {\bf E}\over\partial z} + \delta {\bf M}\cdot
{\partial {\bf B}\over\partial z}.
\end{equation}
We put $\delta {\bf P}=\left(\epsilon -1\right)\delta {\bf E}/4\pi$ and
$\delta {\bf M}=-\partial F/\partial {\bf B}$ where $F=-\epsilon \delta {\bf
E}\cdot \delta {\bf E}/8\pi$ is the magnetic Helmholtz Free Energy of the
plasma in the presence of the wave \citep[see equation 14.1 in][]{landau84}.
Substituting,
\begin{equation}
f_z={\epsilon -1\over 4\pi}\delta {\bf E}\cdot{\partial \delta {\bf E}\over\partial z}+
{\partial\epsilon\over\partial {\bf B}}{\delta {\bf E}\cdot \delta {\bf E}\over 8\pi}
{\partial {\bf B}\over\partial z}={\rho c^2\over 2B^2}{\partial \delta E^2\over\partial z}
-{\rho c^2\delta E^2\over B^3}{\partial {\bf B}\over\partial z}=
{\rho c^2\over 2}{\partial\over\partial z}
\left(\delta E^2\over B^2\right)
\end{equation}
where we have put $\epsilon =1+\omega _{pi}^2/\Omega
_i^2=1+c^2/V_A^2=1+4\pi\rho c^2/B^2$, the only wave property required in this
approach.

\citet{lundin06} give a similar treatment, but based on single particle
dynamics in the {\em external} magnetic field $\left(0,0,B_z\right)$ and the
electric field of the Alfv\'en wave $\left(\delta E_x\exp i\omega
t,0,0\right)$. With
\begin{eqnarray}
i\omega mv_x &=& e\delta E_x+ev_yB_z/c\cr
i\omega mv_y &=& -ev_xB_z/c,
\end{eqnarray}
the Lorentz force on a particle of mass $m$ and charge $e$, we solve to find
\begin{eqnarray}
v_x&=&-{i\omega e\delta E_x/m\over \omega ^2-\Omega ^2}\cr
v_y&=&{\Omega e\delta E_x/m\over \omega ^2-\Omega ^2}.
\end{eqnarray}
The electric field in the $z$-direction which produces the ponderomotive
force is $\delta E_z=-v_x\delta B_y/c$, where $\delta B_y$ is the {\em
internal} magnetic field of the wave, so the total ponderomotive force on a
single particle $\tilde{f}_z$ is given by
\begin{equation}
\tilde{f}_z=-ev_x\delta B_y/c + M{\partial B\over\partial z}
\end{equation}
where $\delta B_y=\left(ic/\omega\right)\partial\delta E/\partial z$ and
$M=exv_y/c=ev_xv_y/i\omega c$ is the magnetic moment of the particle.
Substituting
\begin{equation}
\tilde{f}_z=-{e^2\delta E_x/m\over \omega ^2-\Omega ^2}{\partial\delta E\over\partial z}
-{e^3\over m^2c}{\delta E_x^2\Omega\over\left(\omega ^2 -\Omega ^2\right)^2}{\partial B\over\partial z}
\end{equation}
which for $\omega << \Omega$ reduces to
\begin{equation}
\tilde{f}_z={mc^2\over 2}{\partial\over\partial z}\left(\delta E_x^2\over B^2\right).
\end{equation}

Alternatively, and more generally, \citet{pitaevskii61} and \citet{washimi76}
consider the change induced in $\epsilon$, and hence $F$, the total free
energy, by the passage of a wave. With $F_0$ taken to be the the total plasma
free energy in the absence of wave fields,
\begin{eqnarray}
\delta \epsilon &=&-\left(\delta {\bf r}\cdot\nabla\right)\epsilon -
{\partial\epsilon\over\partial\rho}\delta\rho\nabla\cdot\delta {\bf r}+
{\partial\epsilon\over\partial {\bf B}}\left(\delta {\bf r}\cdot\nabla\right){\bf B}\\
\delta F_0&=&-\left(\delta {\bf r}\cdot\nabla\right)F_0 -
{\partial F_0\over\partial\rho}\delta\rho\nabla\cdot\delta {\bf r}
\end{eqnarray}
where a term in $\partial F_0/\partial {\bf B}=0$ in the absence of waves and
plasma magnetization. Then the perturbed plasma free energy with waves
present, $\delta F$, is
\begin{equation}
\delta F=\int\delta F_0-{\delta E^2\over 8\pi}\delta\epsilon dV=\int\left(\delta {\bf r}\cdot\nabla\right)
\left[-F_0+{\partial F_0\over\partial\rho}\delta\rho -{\delta E^2\over 8\pi}{\partial\epsilon\over\partial
\rho}\delta\rho\right]-{\delta E^2\over 8\pi}
\left[-\left(\delta {\bf r}\cdot\nabla\right)\epsilon +{\partial\epsilon\over\partial {\bf B}}
\left(\delta {\bf r}\cdot\nabla\right){\bf B}\right]dV,
\end{equation}
where terms involving $A\nabla\cdot\delta {\bf r}$ where $A$ is a scalar have
been integrated by parts to give $-\delta {\bf r}\cdot\nabla A$. Equating
$\delta F = -\int {\bf f}\cdot \delta{\bf r} dV$ yields
\begin{eqnarray}
{\bf f} &=& -\nabla\left[{\partial F_0\over\partial\rho}\delta\rho -F_0\right]+{1\over 8\pi}\nabla
\left(\delta E^2\rho {\partial\epsilon\over\partial\rho}\right)-{\delta E^2\over 8\pi}\left[\nabla\epsilon-
{\partial\epsilon\over\partial {\bf B}}\nabla {\bf B}\right]\\
&=&-\nabla\delta P +{1\over 8\pi}\nabla\left(\left(\epsilon -1\right)\delta E^2\right)-{\delta E^2\over 8\pi}
\left[\nabla\epsilon-
{\partial\epsilon\over\partial {\bf B}}\nabla {\bf B}\right].
\end{eqnarray}
For an Alfv\'en wave, $\delta P=0$, $\epsilon =1+4\pi\rho c^2/B^2$ (again,
the only wave property required, like the first derivation above) and
\begin{equation}
{\bf f}={\epsilon -1\over 8\pi}\nabla\delta E^2 -{\rho c^2\over B^3}\nabla {\bf B} \delta E^2
={\rho c^2\over 2}\nabla {\delta E^2\over B^2}.
\end{equation}

\section{Torsional Alfv\'en Waves}
We summarize the magnetic field and velocity perturbations in a torsional
Alfv\'en wave in cylindrical symmetry. The perturbed magnetic field, $\delta
{\bf B}$ is
\begin{equation}
\delta {\bf B}=\nabla\times\left({\bf\xi}\times {\bf B}_z\right)
={B_z\over i\omega}{\partial\delta v_r\over\partial z}\cos m\phi\hat{\bf r}
+{B_z\over i\omega}{\partial\delta v_{\phi}\over\partial z}\sin m\phi\hat{\bf \phi}
-{mB_z\delta v_{\phi}\over i\omega r}\cos m\phi\hat{\bf z}-{B_z\over i\omega r}{\partial\over\partial r}
\left(r\delta v_r\right)\cos m\phi\hat{\bf z}
={B_z\over v_A}\left(\delta v_r\hat{\bf r}+\delta v_{\phi}\hat{\bf
\phi}\right)
\end{equation}
where we take $\nabla\cdot \delta {\rm v}=0$, and $\delta v_r$ independent of
$r$. Hence $\delta v_r = -m \delta v_{\phi}$ and the fluid displacement $\xi
= \delta{\bf v}/i\omega = \left(\delta v_0/i\omega\right)\left(-m\cos
m\phi\hat{\bf r}++\sin m\phi \hat{\bf \phi}\right)\exp\left(i\omega
t+ikz\right)$.

\end{document}